# Leveraging 3D Technologies for Hardware Security: Opportunities and Challenges

Peng Gu, Shuangchen Li, Dylan Stow,
Russell Barnes, Liu Liu, and Yuan Xie
University of California, Santa Barbara
Santa Barbara, CA
{peng_gu, yuanxie}@ece.ucsb.edu

Eren Kursun
Columbia University
New York, NY
ek2925@columbia.edu

## ABSTRACT

3D die stacking and 2.5D interposer design are promising technologies to improve integration density, performance and cost. Current approaches face serious issues in dealing with emerging security challenges such as side channel attacks, hardware trojans, secure IC manufacturing and IP piracy. By utilizing intrinsic characteristics of 2.5D and 3D technologies, we propose novel opportunities in designing secure systems. We present: (i) a 3D architecture for shielding side-channel information; (ii) split fabrication using active interposers; (iii) circuit camouflage on monolithic 3D IC, and (iv) 3D IC-based security processing-in-memory (PIM). Advantages and challenges of these designs are discussed, showing that the new designs can improve existing countermeasures against security threats and further provide new security features.

## 1. INTRODUCTION

Three-dimensional Integration (3D) is a promising technology that allows multiple dies to be vertically integrated in a stack [14]. Successful 3D implementations have been demonstrated [44] in a wide range of scenarios from multi-core processor design and high-bandwidth memory design to network-on-chip (NoC) design. Similarly, interposer based 2.5D integration technology, which interconnects dies by metal routes in the interposer layer, has been explored. Compared with 3D, 2.5D provides benefits of in-package integration with relaxed thermal requirements and design flexibility. An AMD GPU [1] and a Xilinx FPGA [?] have adopted 2.5D integration.

3D and 2.5D integration technologies possess some salient advantages over 2D design [15] such as: (i) providing higher transistor density through die-stacking; (ii) enabling shorter interconnect by partitioning larger dies to smaller ones; (iii) enhancing memory bandwidth by an order of magnitude by eliminating off-chip pin count constraints; (iv) allowing enhanced design flexibility by allowing heterogeneous integration, and (v) offering cost effective alternatives due to improvements in die yield and smaller footprints. While performance, cost and form-factor advantages of 3D integration have been extensively researched in the past, utilizing 3D and 2.5D integration technologies for hardware security enhancement is still an emerging research topic.

In recent years, hardware security has gained significant interest from both industry and academic communities. First of all, various types of attacks aiming at integrated circuits have been increasing rapidly to acquire sensitive information or encrypted keys [45]. Second, there are growing concerns on how to safely manufacture ICs from untrusted manufacturers[36]. While a number of security-aware hardware designs and software algorithms have been proposed to address these challenges, they induce considerable overhead and degrade performance by incorporating security components. With technology advancement, 3D and 2.5D integration solutions can improve the existing countermeasures and further add new security features in protection of emerging threats.

In Section 2 of this paper, we review previous efforts on hardware security enhancement based on 3D integration technology, including split fabrication and modular security integration. Then we introduce our major contributions by presenting four new opportunities using 3D design that can promisingly benefit hardware security: (i) A novel 3D architecture is proposed in Section 3.1 to prevent thermal and power side-channel attacks, taking advantage of 3D IC's intrinsic multi-layer structure and heterogeneous integration. (ii) New optimizations to reduce the cost of split manufacturing through 2.5D-based active interposer design are discussed in Section 3.2. (iii) Monolithic 3D IC is shown in Section 3.3 to potentially support a more effective camouflage circuit to hinder reverse engineering by exploiting the multi-layer structure and efficient interconnects. (iv) In Section 3.4, 3D-enabled processing-in-memory architecture is presented to offset the overhead for memory security with its high-bandwidth memory. In the end of each section, we also discuss the potential design challenges to implement security mechanisms using 3D and 2.5D integration technologies.

## 2. RELATED WORK

Pioneering work has already explored leveraging 3D integration technology for hardware security improvement, including 3D IC-based split fabrication [22, 42] and flexible modular security integration with 3D chips [19, 37].

## 2.1. 3D IC-based Split Fabrication

As the IC manufacturing industry grew into a global business, security and trustworthiness of the components manufactured or designed by third parties have become a serious concern. Split fabrication is proposed to prevent attacks from untrusted foundries, such as hardware trojans, IC overbuilding, and Intellectual Property (IP) piracy [36]. The key idea is to fabricate different parts of the design in different foundries and assemble those parts with a trusted foundry in the end. Even though the foundries are untrusted, with only part of the design, the risk that the attackers can compromise the entire IP is minimized. Digital IC has already demonstrated feasibility for split-foundry processes [18], and fault analysis-based defense methodologies have been proposed to prevent attackers from bypassing the security afforded by straightforward split manufacturing [30].

3D IC-based split fabrication generally follows the original scheme in 2D design. Compared with 2D IC-based split fabrication that partitions the front-end-of-line processing (FEOL, including transistors and lower level metals) and backend-of-line processing (BEOL, including higher-level metals) in different foundries, the 3D IC-based method sends layers to different foundries. Imeson *et al.* [22] examined this method and studied the circuit partitioning problem. Using the proposed k-security partition algorithm and layout randomization, they significantly increased the security level at a low cost. A case study on DES circuit was presented to evaluate the proposed method. Considering the challenges of logic-on-logic 3D stacking, recent studies have also investigated interposer-based 2.5D integration technology. Xie *et al.* [42] proposed a physical design flow for limiting the interposer layer that contains inter-chip connections to be fabricated in a trusted foundry. It provided a security-aware design partition and placement method that generated obfuscated chip layout while achieving a significant security
improvement with tolerable performance/area overhead.

Although 2D IC-based split fabrication works well in most scenarios, 3D integration offers more benefits. First, 3D integration provides further improvement on system performance and cost efficiency. 3D integration lands itself well to split the assembling process. Second, 3D integration provides a much larger design space for security improvement. In 2D IC-based split fabrication, designers can only partition higher metal layers of the chip for security purposes. However, in the 3D scenario, both the transistors and lower metal layers are design knobs. Moreover, in 2D design, the chip can only be split into two parts, but 3D integration provides more flexibility in design with multi-layer die-stacking.

## 2.2. Modular 3D Integration for Security

Sherwood *et al.* [19, 37] proposed an architecture with an extra control plane stacked on top of the existing computation plane. The control plane provides various security mechanisms, and the computation plane contains processing cores. This approach requires minor modifications in the original design such as providing an interface to the control plane. The computation plane is able to function without the control plane. However, after introducing an extra control plane, the critical signals are detoured to the control plane, where they are either encrypted or monitored [7].

Alternatively, by detouring the important signals, the control plane can also bypass the unsecure circuits in the computation plane and replace them with secure design.

This idea takes advantage of the flexibility of 3D stacking, and provides modular security design. It benefits from 3D technologies in the following aspects. The control plane can be integrated in a per-need basis for security-sensitive market segments. Second, thanks to the modular design, a variety of different control plane alternatives can be used with the same computation plane. Third, the 3D stacking offers extremely small latency, large bandwidth, and IO connections. Compared with the off-chip co-processor solution, the performance and power overhead of this modular security design is much lower [38]. Also, this idea could be extended to combine split fabrication [39] and data compression [28].

There are also other techniques combining 3D stacking with hardware security improvements. For example, Bao *et al.* [6] proposed a dual cache design that prevents cache sidechannel attacks. Since the proposed cache design causes performance degradation, 3D cache stacking was introduced to mitigate the performance loss along with other advantages such as the larger bandwidth and shorter access latency.

## 3. NEW OPPORTUNITIES

After re-examining the 3D/2.5D-based security design, we propose four opportunities utilizing these technologies to improve the existing hardware security schemes and show that more security features can be added. We further discuss the potential design challenges for these opportunities. As the attacking techniques become more aggressive, thermal side channel attack emerge, which are difficult for conventional 2D design to prevent. In Section 3.1, we show the 3D technology's potential to tackle various security challenges. Recently, with advancements in processing technology, monolithic 3D integration and active interposers in 2.5D integration technologies are possible, providing more choices for security designs. In Section 3.2 and 3.3, we explain how those techniques benefit security design. Finally in Section 3.4, we show new opportunities in the 3D-supported processing-in-memory architecture.

## 3.1. Shielding Side Channels with 3D Stacking

Side-channel attacks [45] are important security challenges, as they reveal sensitive information about on-chip activities. In cryptographic modules, the cipher hardware is represented as a black box whose internal operations are not observable in theory. However, attackers can bypass the mathematical complexity of the encryption algorithms by extracting and analyzing physical side-channel data such as power dissipation, thermal profiles, electromagnetic radiation, and acoustic traces to determine the secret keys.

Among such attacks, Thermal Side-channel (TSC) attacks have been shown to disclose the activities of key functional blocks and even encryption keys. Built-in thermal sensor data has been used to analyze the task scheduling sequence of encryption algorithms [5].

| proposed tech. | 3D technology's advantage | benefit for security |
|---|---|---|
| Sec. 3.1: shielding side-channel | the die stacking structure | security enhancement |
| Sec. 3.2: split fabrication on active interposers | cost-efficient integration | cost reduction |
| Sec. 3.3: camouflage on monolithic 3D IC | flexible and efficient interconnections | perf./power overhead migration |
| Sec. 3.4: 3D IC-based security PIM | large bandwidth and power efficiency | perf./power overhead migration |

Table 1: Overview of the proposed methods and how they benefit from 3D technology.

Nefarious programs have been shown to use the TSC as a covert communication channel for transferring sensitive information in FPGAs [21] and multicore platforms [27]. By decapsulating an embedded microcontroller and attaching a low-cost thermal sensor, recent studies showed that encryption parameters can be acquired from the temperature profile [20]. The availability of highly sensitive on-chip and off-chip thermal sensors, infrared cameras, and techniques to calculate power consumption from temperature distribution [11] has enhanced the effectiveness of TSC attacks. As a result, side-channel attacks can be performed by using temperature data without measuring power pins of the chip. This improves the accuracy of other attack types such as Differential Power Analysis [25], which relies on power readings.

While a large range of countermeasures has been explored [45], none of these techniques can fully prevent side-channel attacks. In most cases, they aim to make the process more difficult to deter attackers. Even state-of-the-art hardware solutions such as read-proof static coating layers [35] have been shown to be ineffective in preventing such TSC attacks.

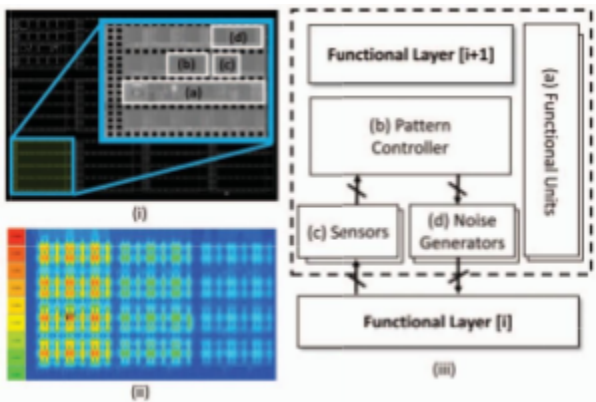

Figure 1: (i) Layout of a Functional Layer. (ii) Patterns Generated by the Pattern Generator Macros. (iii) 3D Side-channel Shielding Hardware Architecture.

The key idea is to protect ICs from thermal side-channel attacks by utilizing intrinsic characteristics of 3D chip integration, as well as proactively using dynamic shielding patterns to conceal critical activities on chip. The proposed technique incorporates targeted strategies to protect from the three attack modes: (i) *Built-in Sensors*: Instead of associating built-in thermal sensors with individual functional blocks, sensors are placed to read out a composite thermal profile of the device and layer blocks. Therefore, the attackers cannot directly associate temperature readings with specific functional blocks while the overall system thresholds are implemented to avoid thermal damage or run-away conditions [17]. (ii) *External Sensors*: Since external thermal sensors can be placed flexibly at various locations by attackers, the effectiveness of the thermal readings can vary. The proposed architecture covers all thermal sensor placement options such that noise injected by security layers will decrease the side-channel leakage of any critical areas. (iii)

*Infrared Thermal Imaging* : The noise generation in the proposed approach conceals the activity patterns of the functional units from infrared cameras and other imaging devices.

We assume the same 3D manufacturing specification as [41]. As demonstrated in Figure 1, this solution includes: (i) A micro-controller unit that dynamically generates complementary activity patterns to prevent side-channel data leakage. Thermal patterns are generated in a randomized, nonrepeating manner such that side-channel attackers cannot extract meaningful information by observing any pattern sequence. (ii) 3D noise generators that run dynamic patterns.

In addition to targeting the individual attack modes separately, the proposed technique works with on-chip power budget and thermal management policies. The power overhead is minimized by intelligently controlling the activity in layers. In side-channel secure mode, the pattern generator macros produce noise according to the security level set by the pattern controllers through the power and thermal constraints. This approach works with on-chip thermal management policies, which will activate appropriate actions in the rare cases where the thermal thresholds are exceeded, causing the whole chip to cool down, without leaking any TSC information.

3D integration provides a number of inherent advantages in side-channel attacks:

*Invasive Attacks*: Historically, removing packaging to reveal device layers has been an effective step in invasive attacks. However, due to the bonding process, 3D layers cannot be removed without harming the normal functionality of the ICs. This protection becomes even more prominent in advanced 3D technologies due to the increased number of thinned device layers [24].

*Semi-Invasive Attacks*: Simple photonic analysis [16]-based attacks require wafer thinning. 3D integration provides protection from such attacks due to the multi-layer stacking that shields the device layers.

*Non-Invasive Attacks*: 3D stacking provides additional protection from non-invasive attacks as device layers and inter-layer bonding materials inherently complicate the radiated side-channel information out of the chip. Thermal profiles of the intermediate layers experience an inherent shielding effect of the layers on top.The proposed 3D architecture could effectively shield the activity patterns in the critical area by the dynamic patterns produced in the noise generators. As the activity level in the functional layer grows, the power consumed by the noise generators will increase accordingly. The challenges are how to minimize the side-channel information leakage under limited power budget, and how to design power-efficient algorithms to maximally reduce side-channel vulnerability while maintaining acceptable power and thermal overhead.

Studies on tradeoff between power consumption and side-channel vulnerability and the relationship between thermal patterns and side-channel information are necessary to optimize runtime management of the proposed 3D architecture.

## 3.2 Cost-aware Hardware Security Using Active Interposers

| $D_0$ : | 500 | 1000 | 1500 | 2000 |
|---|---|---|---|---|
| Passive | 98.5% | 97.0% | 95.5% | 94.1% |
| Active 1% | 98.4% | 96.9% | 95.4% | 93.9% |
| Active 10% | 98.0% | 96.1% | 94.2% | 92.4% |
| Fully-Active | 87.2% | 76.9% | 68.5% | 61.5% |

**Table 2: Interposer yield rates with various active regions calculated by Kannan et al. [23] ($D_0$ is in defects per m$^2$).**

The interposer in 2.5D integration technology connects multiple dies with micro-bumps (μbumps) through interconnecting them with higher-level metal wires. It provides higher degree of integration with less power and smaller latency than PCB or in-package integration, since the μbumps are much smaller than the IO C4 bumps, and the interconnects are basically on-chip metal wires. The interposer technique has already been applied in various products, such as FPGA] and GPU [1]. So far, studies mainly focus on passive interposers, where only high-level metal layers exist. However, 2.5D integration also enables active interposers, which contain both transistors and low-level metal layers. The degree of integration is significantly improved in this approach, while extra costs are incurred due to interposer yield reduction and additional masks. As previous work pointed out [23] in Table 2, as the portion of the active region increases, the yield goes down. For example, a fully-active solution can degrade the yield from 94.1% to 61.5%. Therefore, the number of transistors on the active interposer are typically kept to a minimum.

In this context, split manufacturing techniques (as discussed in Section 2) can be re-examined. With an active interposer, which is already utilized for performance purposes (e.g., high bandwidth memory, HBM [1] integration), an extra 3D layer is not required to split the target die for security purposes. Instead, the die could simply be split into an active interposer and the original die. Now, a cost-aware split fabrication on the active interposer is needed. Previous work developed partition methods either in 3D or passive interposer conditions, where the original designs are equally partitioned into different dies, as shown in Figure 2 (a). However, in split fabrication with an active interposer, the original design is split into both the stacked dies and the active interposer.The challenge is how to minimize the number of transistors placed in the active interposer while meeting the security requirement, as shown in Figure 2 (b). If the number of transistors that are partitioned in the active interposer increases, the yield will decrease, and the cost will rise. In summary, the problem is formulated as minimizing the cost while meeting the security constraints by exploring different partitioning solutions.

## 3.3. Efficient Camouflage IC with Monolithic 3D Technologies

Monolithic three-dimensional (M3D) integration enables revolutionary digital system architectures. Based on metal pitch-scale inter-layer vias (ILV) instead of the large through silicon via (TSV), it provides massive connectivity, leading to ultra-high bandwidth with power consumption reduction. We propose leveraging M3D to enhance security, i.e., as protection from reverse engineering.

The goal of this approach is to provide a camouflage IC design using M3D technology with smaller overhead but better security. The camouflage IC scheme attempts to make the design uniform. Inside a standard cell, the camouflage IC design ensures the layouts of different logic gates to show indistinguishable patterns. Among standard cells, the method tries to fill in the whitespace around macros so that the boundary of the standard cells cannot be recognized. The challenge for conventional camouflage IC design is the performance/area/power overhead. As a result, most chips can only afford partial camouflage of the critical parts. We envision that M3D will improve both inter/intra-standard cell camouflage while reducing lower overhead.

Unlike the current 2.5D and 3D integration using TSVs to connect multiple circuit layers, M3D integration can fabricate a circuit layer directly on top of a previous layer with ILVs, which provides significantly high integration density [40]. Fine-grained M3D integration with CMOS-compatible carbon nanotube FETs or ReRAM [33, 13] at gate-level and circuit-level is also demonstrated.

Reverse Engineering [26] uses imaging recognition techniques to analyze ICs. The images can be obtained by destructive de-layering or nondestructive forms of imaging, after which the schematic can be extracted and analyzed, with the help of pattern recognition and pattern matching [34]. The goal of camouflage IC is to obfuscate the layout (active devices, poly layers, and lower-level metal layers) in order to prevent the reverse engineering. The camouflage IC can be classified into two categories: intraand inter-standard cell camouflage. The intra-standard cell method [31, 9, 8] redesigns the original standard cell. The camouflage standard cell layouts for various gates look identical. It adds a hidden contact layer for each via, which is hard to detect with reverse engineering. The contact layer determines whether the via is connected or just a dummy via. Therefore, with different interconnections, the identical gates provide different functionalities (NAND, NOR, etc.). For the inter-standard cell camouflage, the scheme fills dummy circuits in the empty space among standard cells [10, 12] such that the attackers can not figure out standard cell parts. In M3D context, both the intraand interstandard cell camouflage can be improved, as described below.

**Intra-Standard Cell Camouflage with M3D.** The camouflage standard cell leads to area (4X), delay (1.5X), and power (5X) overhead [31, 32]. Therefore, it is not affordable to replace and camouflage all standard cells. A tradeoff has to be made between the security and the cost. The overhead of camouflaged standard cells, especially for the delay and power, is mainly caused by the extra wiring inside of the standard cell. However, in the M3D, this could be easily solved by fine-grained ILV. Therefore, the overhead is reduced. A series of 3D camouflage IC libraries is required, along with the M3D design flow. M3D also offers a larger design space, which provides more potential for minimizing performance/area/power overhead, and more flexibility to tune the tradeoff between overhead and trust level.

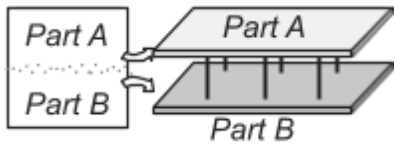

(a) 3D split manufacturing interposer case

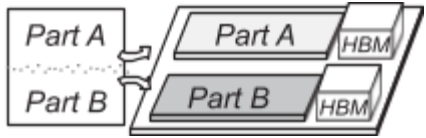

(b) 2.5D split manufacturing

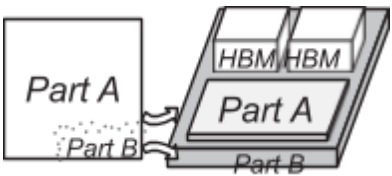

(c) active interposer case

**Figure 2: Previous split fabrication on 3D (a) or passive interposer (b). Cost-aware split fabrication on active interposer (c).**

**Inter-Standard Cell Camouflage in M3D.** The boundary of standard cells which makes reverse engineering more difficult [10] also needs to be blurred. Previous studies proposed filling in the whitespace with dummy cells. In order to hinder the dummy cells, this approach also needs to randomly connect dummy cells to the working cells, which leads to performance degradation. For the M3D structure, it provides higher security with the same overhead. This is because in the 2D design, the rectangle cell has four edges to merge with the dummy cell. For attackers, they only need to recognize 4 edges. However, for 3D monolithic cells, it has 6 faces to merge with dummy cells, which leads to extra difficulties for reverse engineering. Moreover, the 3D partition also makes the attacking challenging, as shown in Figure 3. Multiple partition options increase the search space for reverse engineering.

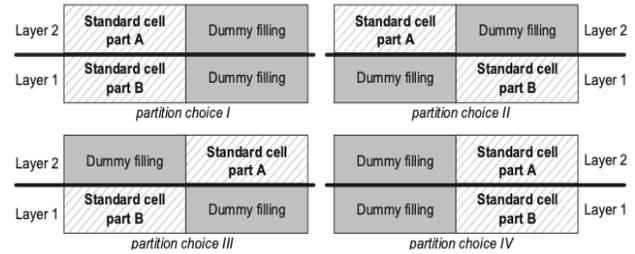

**Figure 3: Inter-Standard Cell Camouflage in M3D.**

## 3.4 Mitigating Security Overhead with 3D PIM Architecture

In order to mitigate the "memory wall", recent studies have focused on a data-centric architecture that tries to bring computing and memory closer. Meanwhile, the processing in-memory (PIM) architecture drew attention [4], due to the significant reduction in data movement overhead. In the early PIM designs, logic and memory were designed on the same die, causing practical challenges to the cost-sensitive memory industry. Therefore, recent PIM studies rely on 3D stacking memory (e.g., hybrid memory cube (HMC) [29]), where the logic and memory are decoupled by die stacking.

The aim is to take advantage of the 3D IC-based PIM's ultra-high bandwidth, massive parallelism, and high energy efficiency to mitigate or release the performance and energy overhead for memory security enhancement. For example, memory authentication leads to 6X bandwidth overhead [43], while 3D-based PIM can potentially provide 80X more bandwidth [2], which totally offsets the performance degradation paid for the memory security. There is an opportunity to integrate secure processors or security protocol accelerators in the logic die in the same stack with the memory dies. Gundu *et al.* [3] already proposed the Security DIMM architecture that leverages PIM for memory security. Their work targeted merkle tree authentication and expected 4.5X speedup, which reduced the overhead from 6X to 2.1X.

## 4. CONCLUSIONS

Security challenges have gained considerable attention from academia and industries as sensitive information and intellectual property may be compromised by the increasing power and number of hardware attacks. Existing countermeasures may induce design overhead, which degrades the performance and adds extra cost. 3D die-stacking and 2.5D interposer design are promising technologies that provide improvements to existing hardware security schemes and include further security features due to the enhancement of integration density, improvement in performance and reduction in cost. This paper summarizes recent studies on hardware security designs based on 3D technology and proposes four opportunities: (i) a 3D architecture for shielding sidechannel information; (ii) split fabrication using active interposers; (iii) circuit camouflage on monolithic 3D IC, and (iii) 3D IC-based security PIM. Advantages and challenges of these designs are discussed and analysed.